# New Security Challenges on Machine Learning Inference Engine: Chip Cloning and Model Reverse Engineering

Shanshi Huang, Xiaochen Peng, Hongwu Jiang, Yandong Luo and Shimeng Yu
School of Electrical and Computer Engineering, Georgia Institute of Technology, Atlanta, GA
Email: shimeng.yu@ece.gatech.edu

*Abstract*—Machine learning inference engine is of great interest to smart edge computing. Compute-in-memory (CIM) architecture has shown significant improvements in throughput and energy efficiency for hardware acceleration. Emerging non-volatile memory technologies offer great potential for instant on and off by dynamic power gating. Inference engine is typically pre-trained by the cloud and then being deployed to the filed. There are new attack models on chip cloning and neural network model reverse engineering. In this paper, we propose countermeasures to the weight cloning and input-output pair attacks. The first strategy is the weight fine-tune to compensate the analog-to-digital converter (ADC) offset for a specific chip instance while inducing significant accuracy drop for cloned chip instances. The second strategy is the weight shuffle to allow accurate propagation of the activations of the neural network only with a key.

## I. INTRODUCTION

In the recent years, machine learning based on deep neural networks (DNNs) has achieved remarkable success in a variety of artificial intelligence applications such as image recognition, language translation, autonomous vehicle, etc. [1]. The success on algorithms also demands the development of a variety of hardware accelerators for their implementations from the could to the edge. Although the mainstream platform for the machine learning research today is GPUs at the cloud due to its computational power and programming flexibility, it is not an ideal choice for edge devices because of its huge power consumption. To this end, hardware accelerators based on application-specific integrated circuit (ASIC) (e.g. Google's TPU [2]) are becoming more and more popular. The primary challenge for machine learning acceleration is the frequent data movements back and forth between the computing units and the memory units. Therefore, compute-in-memory (CIM) [3] is proposed as a promising paradigm for inference acceleration since it overcomes the bottleneck of the conventional von Neumann architecture by emerging the computing units with the memory units. The crossbar nature of the memory array could boost the efficiency of vector-matrix multiplication (VMM) that is widely used in machine learning. The weights of DNNs are mapped as the conductance of the memory array, and the input is loaded in parallel as the voltage to the rows, then the multiplication is done in analog fashion and current summation along columns is used to generate the output. Generally, the CIM architecture could be implemented with SRAM (with modified bit cell [4]) or emerging non-volatile memories (eNVMs [5]). The eNVM based CIM is more attractive to edge devices since they are smaller in size (than SRAM) at the same technology node and will not lose data after power-off, thus do not need to reload DNN model again when power-up. In addition, they have zero leakage which is desired for the battery constrained edge devices. Prototype chips of eNVM based inference engine have been demonstrated in silicon recently, showing impressive energy efficiency 10~100 TOPS/W [6].

However, the non-volatility of DNN model implemented in eNVMs also causes potential threats and vulnerabilities in inference engine. In general, eNVMs suffer from the data privacy problem if the raw data is stored on-chip without encryption [7]. For the CIM architecture, since the analog computation is utilized within the memory array, there is no way to perform the digital encryption. Memory cells have to hold raw data that directly represent the weights of DNN model. This property makes the CIM inference engine under the threats of the chip cloning and the model reverse engineering.

**Asset to be protected:** *The DNN model that is stored in the eNVMs based inference engine is identified as the asset to be protected.* The DNN model owner gathers the training data and labels with great efforts and has to train the network with significant expenses of time and computational resources. If the adversary could easily obtain the DNN model without authorization, he/she could resell it or clone it into another fake chip. First, the training data itself may be private and labeling the training data is a non-trivial effort that typically involves human beings' supervision. Second, the training of DNNs is expensive in terms of computational resources (e.g. cluster of GPUs) and time (e.g. weeks).

**Attack model 1):** *if the adversary could read-out the individual weights that are stored in eNVMs, then the adversary could reprogram (clone) the weights to another chip without going through the expensive training procedure.* In general, micro-probing attack could be used to achieve this goal as a possible way to get around the expensive training procedure. According to [8], it may be challenging to directly probe the eNVM cell without physically damaging the neighboring cells because of its high density. Nevertheless, the adversary could directly probe the digital output from the periphery, e.g. the analog-to-digital converter (ADC). Though VMM computation does not require super high resolution ADC [9], CIM array is typically equipped with high resolution ADC for the write-verify scheme to minimize the variations of the cell conductance. Individual cell's conductance will be read out by such ADC and compared with the reference when initially loading the DNN models on-chip. The ADC output thus becomes a vulnerable spot.

**Attack model 2):** *if the adversary could run the inference engine and gather sufficient input-output pairs, and use them as a new dataset for DNN model reverse engineering.* Attack model 2) has a lower entry bar for adversary than attack model 1) as not everyone has the capability of micro-probing. The threat of input-output pair attack exists as long as the inference engine is supposed to generate the correct output given the input inquiry. Therefore, anyone who obsesses this chip could run the input-output gathering experiments. Obviously, adversary will take efforts to gather sufficient dataset, and it will not eliminate the expensive training procedure (e.g. cluster of GPUs and long training time). However, the adversary is possible to reverse engineer the DNN model without the initial private dataset.

In this paper, we propose a DualSecure scheme to protect the DNN model deployed on the eNVM based inference engine by making it only accessible to an authorized user. To countermeasure attack model 1), *we propose utilizing the process variations of the chip (e.g. ADC offset) to make the DNN model unique for individual chip instance*. The DNN model has to be retrained and adapted to each chip instance's variability anyway. Even the weights are micro-probed and cloned to other chip instances, they would perform inappropriately with low inference accuracy. To countermeasure attack model 2), *we propose the input-output channel shuffle with a key so that only matched shuffle could perform accurate inference*. The key is distributed by the DNN model owner to the authorized user only. Adversary without the key will not be able to generate the correct input-output pairs, thus could not gather the training dataset for model reverse engineering. We experiment our proposed scheme on an 8-layer VGG-like [10] network (VGG-8) model on CIFAR-10 dataset, assuming an eNVM based inference engine manufactured at 40 nm node. The hardware overhead of the proposed DualSecure scheme is also evaluated.

## II. BACKGDOUND OF CIM AND ENVMS

The recent success of artificial intelligence applications such as image recognition is mainly due to the convolutional neural network (CNN), which is a major class of DNNs. Convolution is an important mathematical operation widely used in image processing which extracts the features of images by filters. In the traditional image processing tasks, people manually design the filters' weights in order to extract features such as edge, corner, etc. that they think useful to achieve their goals. The training process of CNN model is to learn the weights of filters iteratively from the training data with the backpropagation that is typically based on stochastic gradient descent method. The inference operation of convolution is essentially the vector-matrix multiplication (VMM). The crossbar nature of eNVMs array is a natural substrate for implementing VMM in a highly parallel manner. As shown in Fig. 1(a), the crossbar array consists of perpendicular rows and columns with the eNVMs cell located at each cross-point. The weights in the filters are mapped as conductance of the eNVMs. The VMM operation is performed as follows: read voltages representing the input feature map are applied to all the rows so that the read voltages are multiplied by the conductance of the eNVMs at each cross-point. The current through each device is summed up along columns. Different columns represent filters for different output channels, who should see the same input thus all the columns work at the same time in parallel. Typically, ADCs are needed at the end of the column to convert the analog current to the digital output so that the subsequent processing such as activation and pooling could be performed in the digital domain.

eNVMs of interests include resistive random access memory (RRAM), phase change memory (PCM), spin-transfer-torque magnetic random access memory (STT-MRAM) and ferroelectric field effect transistor (FeFET). In the recent years, industry has heavily invested in eNVM R&D with commercial processes available, e.g. TSMC 40nm RRAM [11] and Intel 22nm RRAM [12], TSMC 40nm PCM [13], Intel 22nm STT-MRAM [14] and Samsung 28 nm STT-MRAM [15], while doped $HfO_2$ based FeFET technology is also emerging, e.g. Globalfoundries 22nm FeFET [16]. Capitalizing on these progresses, eNVMs based CIM inference engines have also become viable. Most of these industrial processes offer only binary cell, thus we assume 1-bit per cell in this work.

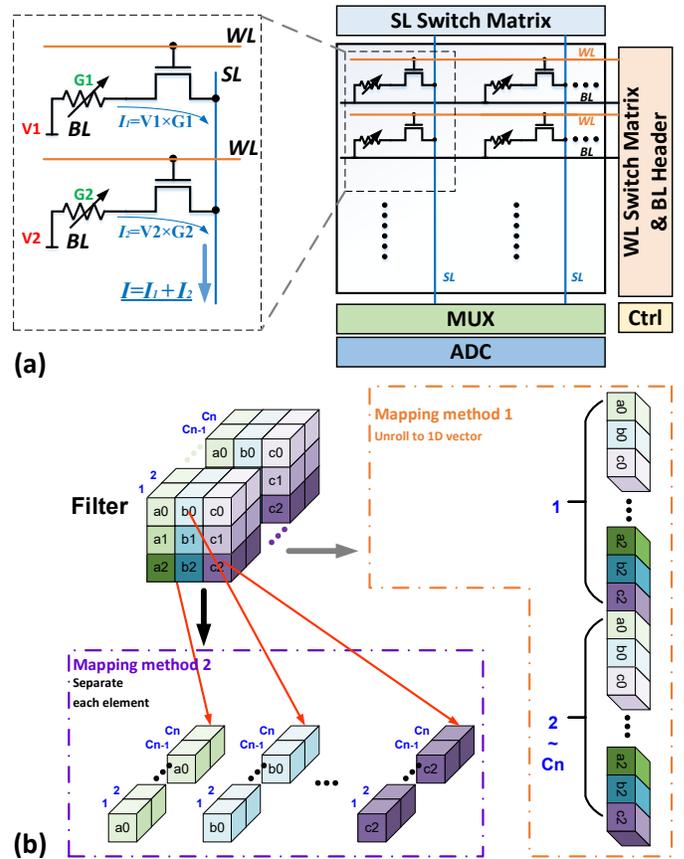

Fig. 1. (a) The crossbar CIM architecture with eNVMs. (b) Two map method to map the convolution filter to the memory column.

Generally, the weights of the convolutional layer is a 4D matrix of size $C_{in} \times C_{out} \times k_1 \times k_2$. $C_{in}$ denotes the channel depth of the 3D input tensor of size $H_{in} \times W_{in} \times C_{in}$ and $C_{out}$ decides the depth of output tensor of size $H_{out} \times W_{out} \times C_{out}$. The weight matrix is applied to the input tensor as a sliding window of size $k_1 \times k_2$. To implement the convolutional layers in memory, there are two kinds of mapping methods to map the model weights to the memory array as shown in Fig. 1(b). Both of them view the weight matrix as $C_{out}$ number of 3D tensor of size $C_{in} \times k_1 \times k_2$. The conventional mapping method stretch the 3D kernel into a long vector and put them in one column of the memory array. Along the row direction, different columns work for different output channels at the same time, so there should be $C_{out}$ number of columns in total for each layer. In deep networks, the depth of the input channel and output channel could be very large that make it hard to fit weights of one layer in a single array considering slow access and extra energy consumption. Array partitioning [17] can be introduced to parallelize the computation into multiple sub-arrays. It is noted that the input data will be reused significantly for convolution as the sliding window moves over the input tensor. To realize the input data reuse practically, a novel mapping methods proposed in [18]. In this method, the weights at different spatial location of each kernel into different sub-kernels. In other words, the 3D kernel is cut into several $C_{in} \times 1 \times 1$ vectors and for each kernel there are $k_1 \times k_2$ of them. These sub-kernels are mapped into different subarrays. Because of the window sliding manner of convolution, sub-kernels of each position will see the input to their neighbor at the window sliding direction. In this case, by passing the used input vectors in the same direction as the kernel "slides over" the input tensor, the input vectors can be reused among the subarrays efficiently. Besides the convolutional layer, there is often another type of layer used in the CNN model which is called fully connected layer. The fully connect layer could be viewed as a special case of $k_1 = k_2 = 1$ and the input tensor of size $1 \times 1 \times C_{in}$, thus could also be mapped to the memory array for calculation.

### III. RELATED WORK

While the neural networks get more and more popular in the real life applications, the security vulnerabilities of it becomes a big problem to worry about. On one side, the neural network itself could be adversarial attacked, poison attacked, etc., that prevent it to work properly. On the other side, the neural network models become a valuable profit since it could bring business advantage. Attacks like model stealing will ease the life of attackers to get a well-trained model or training dataset information and cause profit loss for the model owners.

The adversarial attack currently is the most widely studied attack on DNN-based decision making tasks [19,20]. Adversarial examples, which could be defined as "inputs formed by applying small but intentionally worst-case perturbations to examples from the dataset, such that the perturbed input results in the model outputting an incorrect answer with high confidence"[20], could hurt the functionality of neural network models while are not a problem for human decision. In physical worlds, camera noise, stick on the target object, etc. may lead to adversarial examples causing safety-critical situations like self-driving car. There are a lot of defend methods proposed for this kind of attack [21,22,23], most of which are mainly software technique. Poisoning attack is an attack type that happens during the training of the ML model. Generally, it is done by injecting bad data into the training dataset so that the model achieve good performance [24] or become more fragile during test [25]. Defense technique like [26,27,28] are proposed to protect model from this type to some extent. But still no one for all method found currently.

The functionality aimed attack are general despite the platform on which the model work and could be solved by algorithm effort from software side. So it out of the consideration of our CIM inference engine. The attack types that try to copy or recreate a model are more import for our hardware developer since not all defense method are applicable because the property of the hardware.

Model stealing becomes a problem as the machine learning application move from research field to the business interest. As an example, machine-learning-as-a-service offer well trained model on secured cloud and charge others for access based on pay-per-query. Even though the model could worked as black box, it is proved that the model could be extracted with prediction APIs with limited query[29]. While this problem treat user as potential adversary so that the defense method is to limit the information available to the user or add some tricks on the model to make it hard to be extracted, condition is a little different for the CIM edge device case since the model is local. It hard to count the query number to a local chip and the model works as white box as all the weights are available locally. The model is viewed as part of the chip to be authorized to the user. So in this case, user pay for the chip(model) instead of query.

So in this paper we treat the unauthorized user as adversary. The model could be stolen as chip clone for the CIM edge device. Considering the capability of micro-probing for chip clone, an easier attack method is to extract model from input-output pair since this could be physically achieved by any person and would come back to work as power on since we consider the eNVM-based CIM chip. Previous work [8] have propose a method to prevent input output pair attack by utilize the memristors obsolescence effect. The accuracy will decrease with use so that the edge need to require a secured training dataset from the base station to retrain the model on chip. Only the authorized user will be able to decrypt the training dataset for use. This method will cause big data transfer between cloud and device and need the chip to be equipped with the training functionality, which will be more complicated than the inference only chip. Our method is more focused on inference chip and general to any eNVM devices since we do not rely on any property of the device itself. We didn't consider the non-ideality of the cell, which should be reduced as much as possible for a good device. Although we also apply some training steps on the chip, but it is hybrid so that the chip is inference only after shipped to the user.

### IV. DUALSECURE PROTECTION SCHEME

We propose a DualScheme to protect the eNVM based CIM inference engine. On one side, the DNN model on individual chip is fine-tuned to fit the ADC offset specified to that chip so that it performs poorly on other chips. In this way, the cloned weights becomes useless as ADC offset could not be cloned. On the other side, to prevent input-output pair attack, the weights' input channels are shuffled when programed to the CIM array. A key is needed to shuffle the output from the previous layer to the same order to match the input to the weight matrix of the next layer. The key is only distributed to the authorized user, and will be temporarily stored on volatile on-chip SRAM array so that it could not obtained by the adversary after power-off.

## A. ADC offset variation modeling

ADC is one of the most important periphery circuits for the CIM architecture and has a significant impact on the inference accuracy. Prior work has suggested that even with fine-tuned linear/nonlinear reference levels, ADC quantization (less than full precision of the partial sum of VMM) will introduce a slight accuracy loss [9]. What makes situation worse is the process variations in the actually fabricated chips (e.g. transistor threshold voltage variation, eNVM cell conductance variation). The intrinsic ADC offset makes the partial sum read-out from one memory sub-array different from the correct value after quantization. Though advanced sensing offset cancellation techniques are possible but with significantly increased area [30], CIM architecture has very tight column pitch where ADC size needs to be minimized to maintain the parallelism of the VMM computation.

From the security point of view, we could leverage the process variations to protect CIM inference engine, which is similar to the concept of physical unclonable function (PUF) [31]. Our main goal here is to retrain the DNN model and make it adapt to each chip's variation, thus the model is unique to each chip. We first identify the primary variation source to be utilized. eNVM cell conductance variation is manually controllable (e.g. by aggressive write-verify [32] to tighten the sigma of variation), thus it may not be an ideal source. Instead, ADC offset (for a given circuit topology) is purely defined by manufacturing, thus it is preferred as the source. We could fine-tune the weights to fit the ADC offset pattern of a specific chip so that even the same weights will not perform well for the other chips.

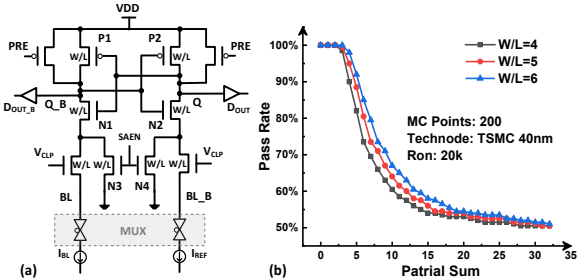

Fig. 2. (a) Latch-based current-mode SA. (b) Sense pass rate for 5-bit ADC.

Generally, there are two types of ADC architecture used in CIM: Flash-ADC and SAR-ADC. The main component in both ADCs that contribute to offset is the sense amplifier (SA). Flash-ADC uses different SAs for different levels and uses encoder to convert the thermometer code to the binary code. SAR-ADC has only one SA but compares for several iterative cycles as a binary tree search to the correct level. We will explore both ADC designs to implement our proposed scheme. Generally, there are two types of SA: voltage mode (VSA) and current mode (CSA). While both SAs could be used in CIM architecture, VSA has less offset than CSA [33]. Since we propose taking advantages of the offset, we choose CSA in this work. Specifically, we use a simple latch based CSA as shown in Fig. 2 (a), to minimize the area of ADC.

In order to obtain the practical ADC offset pattern, we run Monte Carlo simulations in SPICE with TSMC 40nm PDK, which is a technology node that TSMC offers RRAM [11]. We found that the CSA offset is related to both the width of transistors and the column current value. In an analysis of 5-bit ADC, the sense pass rate decreases with increasing partial sum level (or increasing column current) as shown in Fig. 2 (b). If on-state resistance (Ron) of the eNVMs is smaller, the current to be sensed is larger, and the sense pass rate is lower. This trend could be explained as follows: ideally the differential sensing is determined by the relative strength of Ron of the eNVMs and that of the reference cell. When the column current becomes larger as partial sum increases, the voltage drop on the eNVMs becomes less and the conductance of the differential input pair transistors tend to dominate, thus the output is more decided by the intrinsic variation of the differential input pair of the CSA.

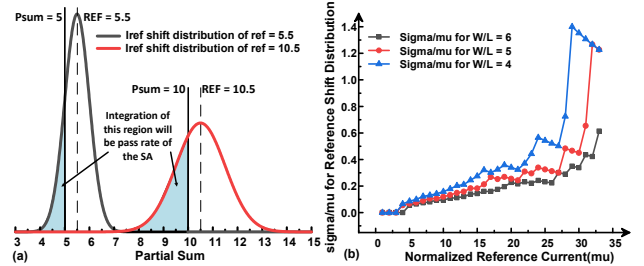

Fig. 3. (a) Sense pass rate to Iref offset conversion (b) Sigma/mu of the Gaussian distribution of Iref offset converted from sense pass rate.

Here the sense pass rate is defined as the probability to sense correctly between the partial sum (Psum) and its nearest reference (Iref). We treat ADC offset as the distance that Iref shifting away from its ideal value, and we assume the Iref distribution follows the Gaussian function. The sense pass rate is interpreted as the cumulative probability of Iref being smaller than the Psum as the blue shade area shown in Fig. 3 (a). We could reversely obtain the sigma of Iref shift from the sense pass rate based on the cumulative probability function of Gaussian. For each SA, since the offset is caused by manufacturing, the Iref shift is a spatial variation and it will be stationary over time. For a 5-bit Flash ADC, there are 31 different SAs which may have different shifts from each other. For the SAR ADC, since the same SA is always used, for each level, the Iref should be shifted to the same direction. Based on the observation that the sigma over mu ratio is increased with Iref (Fig. 3(b)), we scale up the absolute shift value (normalized to Iref) by the ratio between them.

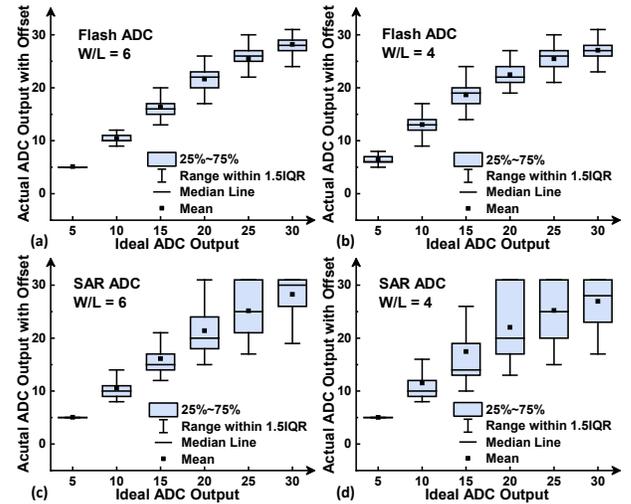

Fig. 4. Simulated ADC output with offset sampled from the Iref distribution based on the sense pass rate for different W/L for Flash-ADC and SAR-ADC.

From the DNN model's perspective, the Iref shift could be viewed as partial sum quantization bias. For Flash-ADC, since each shift is independent, somehow these random shifts could compensate each other. But for SAR-ADC, the bias favors in one direction. As shown in Fig. 4. SAR-ADC has bigger variation when sensing the same partial sum than Flash-ADC. Increasing transistor size (e.g. W/L) will reduce the mismatch, thus increasing the sense pass rate. In this work, as we embrace the process variations, SAR-ADC is a better choice for improving security as we will retrain the DNN model anyway.

We now discuss the DNNs model retrain methods. A software-hardware hybrid method is used to fine-tune the weights after manufacturing for each chip, which means the feedforward propagation (inference) is done on-chip and the rest steps (error calculation, gradient calculation, and weight update) are done off-chip by software. For a specific chip, we will run the inference in hardware using its own ADC offset pattern with input images (selected from one epoch of 50,000 CIFAR-10 images as the retrain dataset), then we will compare the prediction of the inference with the ideal label for the loss function. With the estimated loss, we will run the backpropagation to calculate the weight update in software. Then we reprogram the conductance of the eNVMs with write-verify to the new weights. The weights of the entire neural network is susceptible to retrain. Typically, 1~2 retrain epoch is needed to achieve a reasonably high accuracy for a specific chip. In our evaluation, for the inference, the input and weights are converted to binary sequences so that the VMM result of them will be the normalized partial sum current. Then this value will be mapped to the ADC output with offset by one group of sampled reference currents from the distribution as in Fig. 3(a). Finally, the actual output with offset will be converted back to decimal value as the output feature map and saved for error and gradient calculation. The backpropagation and weight update are all directly use floating-point calculation as in software.

*B. Output-Input channel shuffle*

Generally, in order to protect the data, the data should be encrypted so that even the unauthorized user read it out, he/she is not able to be decrypt it without the key. However, for the eNVM based CIM inference engine, if we encrypt the weights that are stored in the memory array, we have to read out individual weight and decrypt it with the key for digital calculation. Essentially, the analog computation within the memory array is compromised, as it is no longer in-memory computing but more like near-memory computing. In addition, it is slow and energy-inefficient to read the weights row-by-row and decrypt them at the edge of the array for the subsequent digital multiply-and-accumulate operations.

Since it is difficult to directly encrypt the weights, another method for obfuscating the unauthorized user is to play tricks on the input. The inference of DNN model is typically conducted layer by layer, it is important for each channel to receive the correct input from the previous layer's output. This requires that the input to the CIM memory array's rows always being matched. Even if the calculation of the previous layer is correct, as the following layer sees the incorrect input feature map paired with the weights, all the subsequent calculation will be incorrect. Hence, we propose shuffling the weights when programmed into the CIM memory array. Meanwhile the input feature map has to be shuffled in the same way so that the channel of the input and the weight matrix match. What we need to protect is the shuffle scheme, without which the adversary could not perform VMM correctly if using the weights the stored on-chip. This shuffle scheme could be saved on cloud as a key and each time could be downloaded upon the user or chip's successful authentication.

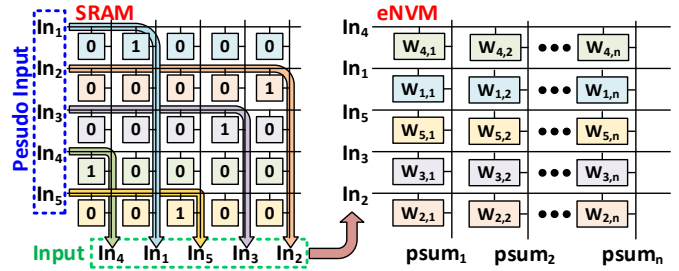

Fig. 5. Architecture of the CIM-based Output-Input channel shuffle method. SRAM array stores shuffle pattern, and eNVM array stores weight pattern.

To minimize the impact on inference latency, we design the shuffle in the CIM fashion. In particular, we propose using a SRAM-based CIM array to perform the shuffle. Each column of the SRAM array stores one hot code to present the shuffle scheme for certain input channel. The output of the SRAM array will be the shuffled channel that matches the weight channel order as shown in Fig. 5. The pseudo-input (from the previous layer's output) is loaded into the SRAM array's row. Given the shuffle pattern, the column output from SRAM array will be the real-input for the eNVMs array where the weights are stored.

We choose to use SRAM to store shuffle pattern based on the following considerations. First, we do not wish to store the shuffle scheme on-chip during power-off. Second, each time when starting inference, we need to write the shuffle pattern (one hot code) back into the CIM array, it takes much less time and power to write SRAM than writing eNVMs. Since each column is one hot coded, the column output will be either 1 or 0. The CIM array just needs 1-bit SA instead of multi-bit ADC. The modifications to this SRAM array is to replace decoder with switch matrix as we have to open all the rows for parallel computation. The disadvantage of using SRAM over eNVMs is that the area is larger. We will explore the associated hardware overhead in the next section.

The question is how safe will this method be. For naïve consideration, this method could be attacked by the brute and force method. Let's check the effort needed to attack an array of size N. The probability to find the exact key is obviously small when N is large. However, the network is not need all the input-output channel match to work. Assume n out of N total digits matched between the real key and a random key, we want to find the probability as a function of n. It is not easy to find the exact number of it. However we know that this value is smaller than the probability of at least n digits match and further smaller than the equation 1. So that we have an upper bound of the probability to find a key have n digits matched with the real key.

$$upper\ bound = \left[\binom{N}{n} \cdot (N-n)!\right] / N!$$

It is obvious that the bigger the array is, the harder to break the key for shuffling. Since the length of the key is related to input channel depth of the layer, dependents on the model structure, the input channel depth may not be high enough to get a strong key. So in order to further increase the security level, we could insert some fake rows into the weight array. These fake row should not change the partial sum during propagation of the data. This could be down by making the input of the row to be zero, which means we have to insert zero into the input of the weight. This could be done by inset zero column in to the shuffle array as shown in Fig. 6. Since the zero insertion don't need to care about the order so it does not mean to increase N directly. The upper bound is changed to

$$upper\ bound = \left[\binom{N}{n} \cdot (M-n)!/k!\right] / \left[\binom{M}{k} \cdot (N)!\right]$$

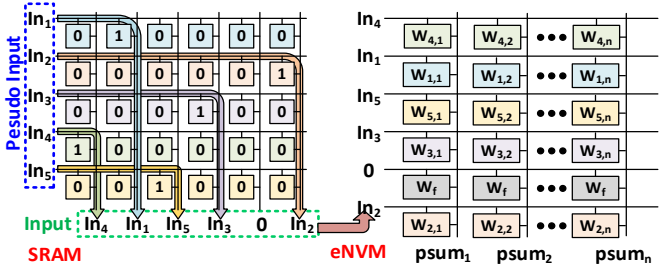

Fig. 6. Architecture of the CIM-based Output-Input channel shuffle with zero insertion method. SRAM array stores shuffle pattern, and eNVM array stores weight pattern with some fake weights inserted.

## V. EVALUATION RESULTS

Our evaluation on the algorithm-level performance is done by the PyTorch platform [34], and the hardware performance is estimated by the NeuroSim framework for CIM [35]. We evaluate the CIFAR-10 image classification on a VGG-8 model. The basic setting for inference precision is that activations (input) are 8-bit and the weights are 2-8 bit. Our software baseline accuracy is ~92%. Binary RRAM with Ron=20kΩ is used as the eNVMs technology in this work.

### A. Retrain for ADC offset and immunity for weight cloning

We first test the inference accuracy with the ADC offset on the original model that is pre-trained by software (e.g. in the cloud), which means we have not considered any retrain to specific chip at the edge. It is seen from Fig. 7 that for the same W/L, SAR-ADC could introduce more accuracy drop than Flash-ADC although the SA sense pass rate is kept at the same. This is in agreement with discussions in Section III on the possible compensation of the quantized partial sum for Flash-ADC. As W/L decreases, the accuracy drops more due to larger mismatch and lower sense pass rate.

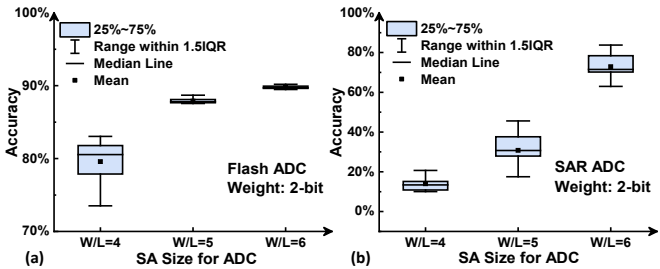

Fig. 7 Inference accuracy distribution of software trained model (VGG-8 on CIFAR-10) with ADC offset.

Fig. 8 shows the retrain curve of Flash-ADC and SAR-ADC with different W/L (thus different offset) for one specific chip. Flash-ADC could be retrained to recover the accuracy, however Flash-ADC has a small initial accuracy drop as it may not be preferred for security purpose as discussed above. We also vary the weight precision of SAR-ADC and evaluate its impact on retrain performance. It is seen that as the W/L decreases, it will be more difficult to retrain the model to recover the accuracy under process variations. For the same W/L, higher weight precision will be more robust to the variations and thus sees less accuracy drop and easier retrain recovery. To balance security and performance, SAR-ADC with precisely chosen W/L should be used.

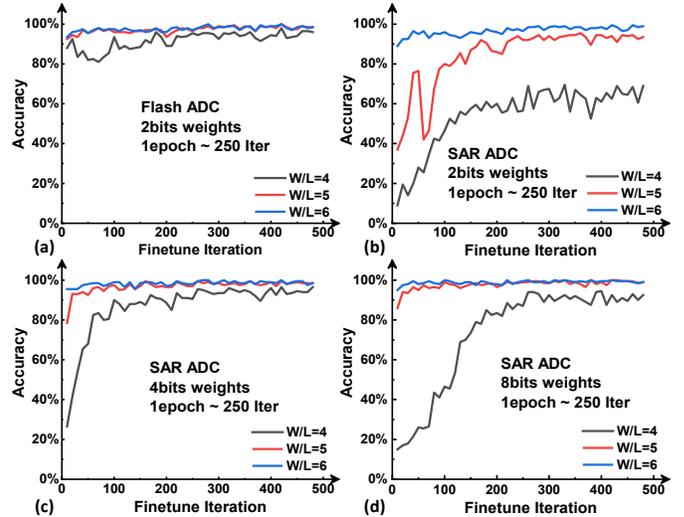

Fig. 8 Retrain curve of different ADC type and weight precision for a specific chip with certain ADC offset.

Now we assume Attack model 1) occurs on this specific chip. The adversary is able to micro-probe each RRAM cell's conductance accurately and is able to reprogram the weights to another chip accurately by write-verify. We evaluate the effectiveness of our countermeasure scheme by exploiting the ADC offset, which the adversary has no control on another chip. We applied the retrained weights of this specific chip to other chips that have different ADC offset patterns. Based on the observation from the retrain curve of SAR-ADC in Fig. 8, we choose W/L=6 for 2-bit weight and W/L=5 for 4/8-bit weights as the design specs for the chip. First, we run several retrain tests to show that the recovery of accuracy to ~91% is not a one-time coincidence (Fig. 9(a)). Then, we apply the retrained model to other chips assuming the weights are cloned to other chips. As shown in Fig. 9(b), the retrained model does result in relatively low accuracy 40%~70% when applied to other chips that may have different ADC offset patterns. Essentially we utilize retrain to enhance the accuracy for this specific chip against the process variations, while rely on the process variations to hamper the accuracy for other chips. Adversary could just clone the weights but could not clone the same performance to other chip instances. Nevertheless, by micro-probing the adversary obtains the weights in the DNN model and can still resell the model itself in the software format (though the model may not be generalized as high accuracy is only associated with this ADC offset pattern specific to this chip). Therefore, this attack become invaluable.

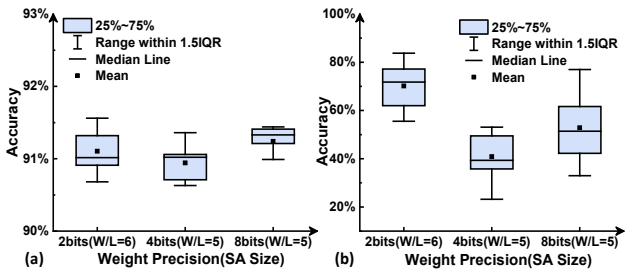

Fig. 9 (a) Inference accuracy distribution of retrained model with ADC offset on this specific chip. (b) Inference accuracy distribution of retrained model applied on other chips.

### B. Output-Input channel shuffle

First we check the hardware overhead introduced by the shuffle unit. We set the RRAM subarray size to 128×128 and SAR-ADC as 5-bit. The detailed hardware configurations are shown in Table I and the performance is evaluated at 40 nm node by NeuroSim. It is seen that the SRAM array (shuffle array) could be read in parallel, and it is much faster than the VMM computation in the RRAM array. Thus, shuffle array could be shared by several weight array without introducing significant delay overhead. It is noted that the SRAM array is larger than the RRAM array. When the weight precision is low, the overhead is quite large if all the layers in the neural network are equipped with the shuffle array (e.g. ~50% for 2-bit weight). Fortunately, as shown later it is not necessary to shuffle every layer to ensure security, thus the total overhead could be maintained to an acceptable level.

| Component | Spec. | Area (mm²) | Latency (ns) | Energy (pJ/op) |
|---|---|---|---|---|
| RRAM array | 128×128 (1 bit per cell) | 855.436 | 32.027 | 55.2594 |
| wlSwitchMatrix | | 350.644 | | |
| slSwitchMatrix | | 236.357 | | |
| SAR-ADC | 5-bit (×16) | 5221.339 | 112.945 | 85.174 |
| ShiftAdd | 14-bit (×16) | 1012.92 | 0.84 | 10.07 |
| Total | | **7029.7** | **145.813** | **150.5** |
| SRAM array | 128×128 | 5994.12 | 1.251 | 3.3 |
| wlSwitchMatrix | | 1293.773 | | 2.825 |
| precharger | | 511.527 | | 5.176 |
| WriteDriver | | 511.527 | 0 | 0 |
| SenseAmp | 1-bit (×128) | 334.705 | 0.12 | 23.552 |
| Total | | **8645.652** | **1.371** | **34.853** |

Table I. Hardware configurations and performance for RRAM CIM array (weight array) vs. SRAM CIM array (shuffle array).

Now we assume Attack model 2) occurs as adversary tries to run this inference engine to gather the input-output pairs. We evaluate the effectiveness of our countermeasure scheme using channel shuffle as follows. First, we assume the weights of all the convolutional layers are shuffled except the first one since the channel depth of the input images is just 3 (for RGB channels) thus it is easy to guess with limited shuffle combinations. Furthermore, we do not add shuffle to the fully connected (FC) layer since the weight sharing is low for FC layer so the penalty of input shuffle will be relatively high. Since the adversary does not have prior knowledge about the key used for shuffle, we assume that a random key is used by the adversary. Fig. 10 shows the inference accuracy under randomly generated key for (a) different number of shuffle layers in the network and (b) one layer but in different locations of the network. It is seen that even one shuffle layer results in very low (<20%) inference accuracy. Therefore, we will use one-layer shuffle scheme and evaluate its hardware overhead. Fig. 11 shows the entire chip overhead in terms of (a) area and (b) energy. Adding shuffle to shallow layers (such as conv2) introduces small area overhead than deeper layers since the channel depths are smaller for shallow layers. Lower weight precision (thus smaller RRAM array) will make the SRAM shuffle array takes more portion of area, thus larger area overhead. The energy overhead is comparable regardless where the shuffle layer is, because channel depths increase while size of input feature map decreases as layer goes deeper. This means that there will be less shuffle arrays working for more cycles at shallow layers while more arrays working for less cycles at deep layers.

Considering the shuffle operation consumes additional power, it is undesired to shuffle all the bits when the weight precision is high (e.g. 8-bit). By observing the accuracy drop trend in Fig. 12, we found shuffling the 1st and 2nd most significant bit (MSB) is critical for reducing accuracy while the energy consumption will increase linearly as increasing the number of bits to be shuffled. It is noted that in hardware design we map different significant bits into different RRAM arrays and perform shift and add to obtain the final sum. Therefore, adding shuffle to some MSB arrays is possible.

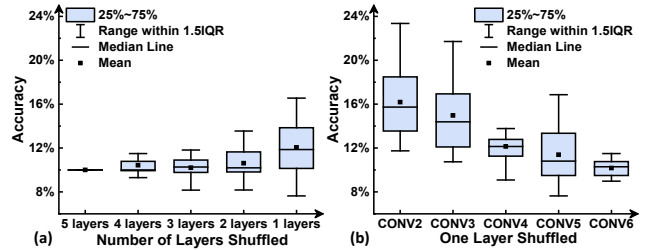

Fig. 10. (a) Inference accuracy distribution vs. different number of layers with a random key. (b) Inference accuracy distribution of one layer but different locations in the network with a random key.

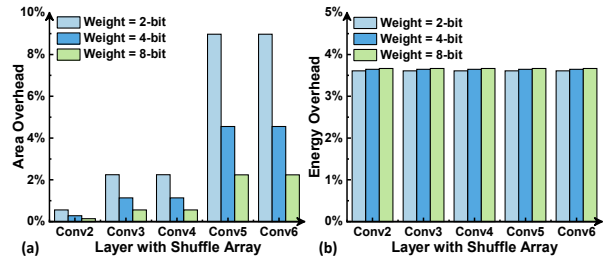

Fig. 11. (a) Area overhead with respect to the location of shuffle array. (b) Energy overhead with respect to the location of shuffle array

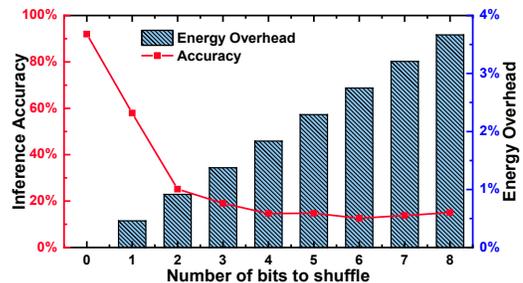

Fig. 12. Tradeoff between inference accuracy and energy overhead based on the number of bits shift from MSB to LSB.

Then, let's check the effect of number of matched digits on the accuracy. From Fig. 13, we found that with the increase number of matched digits, the accuracy degradation will decrease. If half of the digits N/2=64 is matched, the accuracy degradation is not enough for security consideration. However, from the equation 1, we know that the probability boundary of n digits matched out of N=128 is shown as Fig.14. Even for half of position match, it is a very small probability for brute and force attack.

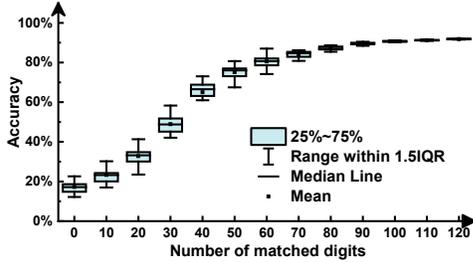

Fig. 13 Accuracy under different number of digits match between real key and random key.

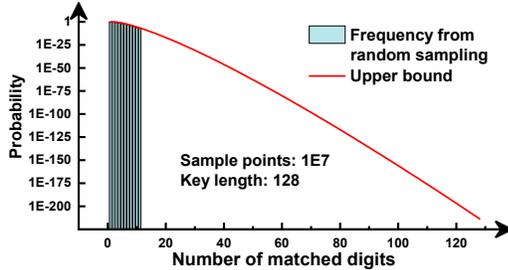

Fig. 14 Probability upper bound for different number of digits match between real key and random key. An experiment is run to generate random key and check the frequency of matched digits with the real key. It could be see that it match the upper bound result when n is small. For big n, since the sample points is limited, no result is viewed.

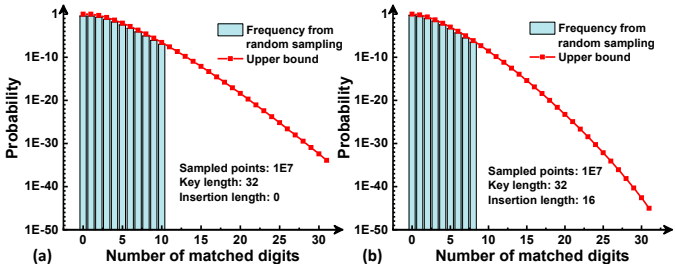

Fig.15 Effect of zero insertion on the probability of number of matched digits

We assume a small Cin layer which has N=32 and check how will inserting 16 fake zeros affect the probability. For Fig. 15 we could see that with zero inserted, probability of matched digits decrease faster than the original case. The problem of this method is that it introduce hardware overhead one the shuffle array. And people may argue that normally the input channel depth may get bigger as the model get deeper. If the shallow layer is not deep enough, then guard one that is deeper. The point is that the model generally is defined without knowing the hardware setting. It may be hard to find the proper depth to balance the hardware overhead and the security power. With zero insertion, we could insert proper number of zeros to make array size desired while increasing the security level.

## VI. CONCLUSION

In this paper, new threats of eNVM based machine learning inference engine are identified including the weight cloning and model reverse engineering. We proposed a DualSecure scheme against these vulnerabilities. Our evaluation results show that by utilizing the process variations (i.e., ADC offset), the DNN model could be retrained to maintain high accuracy on each individual chip while their performance will significantly degrades on other chips even with cloned weights. While this method could effectively countermeasure chip cloning, the model reverse engineering by input-output pair attack is still possible. We further propose to add an SRAM-based CIM shuffle array so that a key is needed to match the input channel with the weight matrix. By applying the shuffle scheme on just one shallow layer and the first two MSB bits of the weights in the network, it could achieve high security with relatively low overhead for area, latency and energy. We also check the security strength of this shuffled array method and proposed zero insertion method to balance the software define model depth and the security level. This work raises the new research directions in the machine learning hardware security beyond the software adversarial attack.

## REFERENCES

[1] L. Yann, et al., "Deep learning," in *Nature 521 (7553): 436-444,* 2015.

[2] N. P. Jouppi, et al., "In-datacenter performance analysis of a tensor processing unit," in *ACM/IEEE International Symposium on Computer Architecture (ISCA)* 2017.

[3] D. Ielmini, et al., "In-memory computing with resistive switching devices," in *Nature Electronics, 1, no.6: 333,* 2018.

[4] W.-S. Khwa, et al., "A 65nm 4Kb algorithm-dependent computing-in-memory SRAM unit-macro with 2.3 ns and 55.8 TOPS/W fully parallel product-sum operation for binary DNN edge processors," in *IEEE International Solid-State Circuits Conference (ISSCC)*, 2018.

[5] S. Yu, "Neuro-inspired computing with emerging non-volatile memory," in *Proc. IEEE*, *vol. 106, no. 2, pp. 260-285,* 2018

[6] C.-X. Xue, et al., "A 1Mb multibit ReRAM computing-in-memory macro with 14.6 ns parallel MAC computing time for CNN based AI edge processors," in *IEEE International Solid-State Circuits Conference-(ISSCC)*, 2019.

[7] S. Ghosh, et al., "Security and privacy threats to on-chip non-volatile memories and countermeasures," *IEEE/ACM International Conference on Computer-Aided Design (ICCAD)*, 2016.

[8] C. Yang, et al "Thwarting replication attack against memristor-based neuromorphic computing system" in *IEEE Transactions on Computer-Aided Design of Integrated Circuits and Systems,* 2019.

[9] X. Sun, et al., "XNOR-RRAM: A scalable and parallel resistive synaptic architecture for binary neural networks," in *IEEE Design, Automation & Test in Europe Conference (DATE), pp. 1423-1428,* 2018.

[10] K. Simonyan, et al., "Very deep convolutional networks for large-scale image recognition," *in International Conference on Learning Representations (ICLR),* 2015.

[11] C. C. Chou, et al., "An N40 256K× 44 embedded RRAM macro with SL-precharge SA and low-voltage current limiter to improve read and write performance," *IEEE International Solid-State Circuits Conference (ISSCC), pp. 478-480,* 2018.

[12] P. Jain, et al., "A 3.6 Mb 10.1 Mb/mm 2 embedded non-volatile ReRAM macro in 22nm FinFET technology with adaptive forming/set/reset schemes yielding down to 0.5 V with sensing time of 5ns at 0.7 V," *IEEE International Solid-State Circuits Conference (ISSCC), pp. 212-214,* 2019.

[13] J.-Y. Wu, et al., "A 40nm low-power logic compatible phase change memory technology," *IEEE International Electron Devices Meeting (IEDM)*, 2018.

[14] L. Wei, et al., "A 7Mb STT-MRAM in 22FFL FinFET technology with 4ns read sensing time at 0.9 V using write-verify-write scheme and offset-cancellation sensing technique," *IEEE International Solid-State Circuits Conference (ISSCC),* 2019.

[15] Y.-K. Lee, et al., "Embedded STT-MRAM in 28-nm FDSOI logic process for industrial MCU/IoT application," *IEEE Symposium on VLSI Technology,* 2018.

[16] S. Dunkel, et al. "A FeFET based super-low-power ultra-fast embedded NVM technology for 22nm FDSOI and beyond," *IEEE International Electron Devices Meeting (IEDM),* 2017.

[17] P.-Y. Chen, et al., "Partition SRAM and RRAM based synaptic arrays for neuro-inspired computing*," IEEE Int. Symp. Circuits Syst. (ISCAS),* 2016.

[18] X. Peng, et al., "Optimizing Weight Mapping and Data Flow for Convolutional Neural Networks on RRAM Based Processing-In-Memory Architecture." *2019 IEEE International Symposium on Circuits and Systems (ISCAS). IEEE, 2019.*

[19] C. Szegedy, et al., "Intriguing properties of neural networks." *International Conference on Learning Representations (ICLR). 2014.*

[20] I Goodfellow, et al., "Explaining and harnessing adversarial examples." *International Conference on Learning Representations (ICLR). 2014*

[21] F. Liao, et al. "Defense against adversarial attacks using high-level representation guided denoiser." *Proceedings of the IEEE Conference on Computer Vision and Pattern Recognition(CVPR). 2018.*

[22] C. Xie, et al. "Feature denoising for improving adversarial robustness." *Proceedings of the IEEE Conference on Computer Vision and Pattern Recognition(CVPR). 2019.*

[23] J. Cohen, et al. "Certified adversarial robustness via randomized smoothing*." arXiv preprint arXiv:1902.02918 (2019).*

[24] S, Jacob, et al., "Certified defenses for data poisoning attacks." Advances in neural information processing systems. 2017.

[25] X. Chen, et al. "Targeted backdoor attacks on deep learning systems using data poisoning." arXiv preprint arXiv:1712.05526 (2017).

[26] M, Jagielski, et al. "Manipulating machine learning: Poisoning attacks and countermeasures for regression learning." *2018 IEEE Symposium on Security and Privacy (SP), 2018.*

[27] Y. Gao, et al. "STRIP: A Defence Against Trojan Attacks on Deep Neural Networks." *arXiv preprint arXiv:1902.06531 (2019).*

[28] W. Guo, et al. "Tabor: A highly accurate approach to inspecting and restoring trojan backdoors in ai systems*." arXiv preprint arXiv:1908.01763 (2019).*

[29] N. Papernot, et al. "Practical black-box attacks against machine learning." *Proceedings of the 2017 ACM on Asia conference on computer and communications security, 2017.*

[30] C.-P. Lo, et al., "Embedded 2Mb ReRAM macro with 2.6 ns read access time using dynamic-trip-point-mismatch sampling current-mode sense amplifier for IoE applications," *IEEE Symposium on VLSI Circuits,* 2017.

[31] R. Liu, et al., "A highly reliable and tamper-resistant RRAM PUF: design and experimental validation," *IEEE International Symposium on Hardware-Oriented Security and Trust (HOST),* 2016.

[32] L. Gao, et al., "Programming protocol optimization for analog weight tuning in resistive memories," *IEEE Electron Device Lett., vol. 36, no. 11, pp. 1157–1159,* 2015.

[33] S.-H. Woo, et al. "Offset voltage estimation model for latch-type sense amplifiers," *IET Circuits, Devices & Systems 4.6: 503-513*, 2010.

[34] A. Paszke, et al. "PyTorch: An imperative style, high-performance deep learning library," *Advances in Neural Information Processing Systems*, 2019.

[35] P.-Y. Chen et al., "NeuroSim: A circuit-level macro model for benchmarking neuro-inspired architectures in online learning," in *IEEE Transactions on Computer-Aided Design of Integrated Circuits and Systems,* vol. 37, no. 12, pp. 3067-3080, 2018.